# 基于 CNN-SAEDN-Res 的短期电力负荷预测方法

崔　杨[1]，朱　晗[1]，王议坚[1]，张　璐[2]，李　扬[1]

(1. 东北电力大学 现代电力系统仿真控制与绿色电能新技术教育部重点实验室，吉林 吉林 132012；2. 中国农业大学 信息与电气工程学院，北京 100083)

摘要：基于深度学习的序列模型难以处理混有非时序因素的负荷数据，这导致预测精度不足。提出一种基于卷积神经网络（CNN）、自注意力编码解码网络（SAEDN）和残差优化（Res）的短期电力负荷预测方法。特征提取模块由二维卷积神经网络组成，用于挖掘数据间的局部相关性，获取高维数据特征。初始负荷预测模块由自注意力编码解码网络和前馈神经网络构成，利用自注意力机制对高维特征进行自注意力编码，获取数据间的全局相关性，这使模型能根据数据间的耦合关系保留混有非时序因素数据中的重要信息，通过解码模块进行自注意力解码，并利用前馈神经网络回归初始负荷。引入残差机制构建负荷优化模块，生成负荷残差值，优化初始负荷。仿真结果表明，所提方法在预测精度和预测稳定性方面具有优势。

## 0　引言

提高负荷预测的准确度和可靠度有利于电网经济、安全运行。目前，短期负荷预测方法主要分为两大类，即基于时序分析的统计学方法和基于数据驱动的机器学习方法。其中，统计学方法于上世纪 70 年代被引入，该方法在不考虑相关因素对负荷影响的前提下，通过分析历史负荷数据来探索负荷的周期性变化[1-2]。虽然能有效处理平稳的负荷时间序列变化，但对于天气因素骤变、设备故障检修、特殊活动等所带来的非线性负荷变化无法有效处理，即非线性学习能力差。

后来，随着智能计算的发展，基于数据驱动的机器学习方法被应用到负荷预测领域[3]。该类方法通过对包含负荷、气象、日期等在内的各种数据进行处理，提取并选择能够充分反映未来负荷变化的特征，使负荷变化与外界因素相关联，从而实现高精度预测。

当前，应用于负荷预测的机器学习方法主要包括传统的机器学习方法[4-5]和深度学习模型。传统的机器学习方法需要先开展特征工程工作，对数据进行深度分析后才能建立模型进行学习，否则预测效果便会不理想。相比之下，深度学习模型可以直接将数据传递到网络，利用其强大的非线性表达能力挖掘特征，建立负荷与各类因素之间的关联性，实现高精度预测。

在深度学习模型中，最具代表性的为卷积神经网络（convolutional neural network，CNN）和序列模型（循环神经网络及其变体）[6-7]，不少研究者基于此提出了负荷预测模型，如利用长短期记忆（long-short term memory, LSTM）模型[8]、门控循环单元（gate recurrent unit, GRU）模型[9]挖掘负荷数据的时序性特点。有些工作考虑到序列模型不能挖掘特征在高维空间的联系，因此将卷积神经网络和序列模型结合起来组成混合神经网络，由卷积神经网络提取高维特征，再交由序列模型处理[10-13]。但需要注意的是，序列模型仅适合处理时间序列数据。如果输入数据中混入日期、气象等非时序因素，模型便无法发挥其特性，去挑选重要信息，来让之前的输入对未来的输出产生正确的影响。换言之，无法像时间序列那样建立前后信息之间的耦合关 系，对预测结果产生积极的推动作用。

因此，本文提出了一种基于卷积神经网络、自注意力[14]编码解码网络（self-attention encod- er-decoder network, SAEDN）以及残差优化（residual-refinement, Res）的短期电力负荷预测方法。该方法可不受数据类型的限制，利用自注意力机制建立特征中任意信息之间的耦合关系，并通过耦合程度自适应的选择保留重要特征和遗忘不重要特征。此外，引入残差机制，在负荷回归之后添加负荷优化模块优化初始负荷，进一步提升模型预

收稿日期：2022-12-21；修回日期：2023-06-07

基金项目：现代电力系统仿真控制与绿色电能新技术教育部重点实验室开放课题(MPSS2021-09)；吉林省自然科学基金资助项目（YDZJ202101ZYTS149）

Project supported by the Open Fund in the Key Laboratory of Modern Power System Simulation and Control & Renewable Energy Technology，Ministry of Education（MPSS2021-09）and the Natural Science Foundation of Jilin Province（YDZJ202101ZYTS149）

测精度。

# 1 基于 CNN-SAEDN-Res 的负荷预测方法

## 1.1 模型概述

如图 1 所示，本文所提出的负荷预测模型包含特征提取模块、初始负荷预测模块、负荷优化模块。预测流程为：首先，通过特征提取模块提取特征，构建数据之间的邻近（局部）特征关系。然后，对提取到的特征进行二维空间位置编码，为其赋予空间位置信息。接着，将编码后的特征送入到自注意力编码解码网络中进行编码和解码，再经过前馈神经网络（feedforward neural network, FFN）回归日负荷数据。最后，将初始预测负荷与原始数据一同送入到优化模块中，优化初始预测负荷。

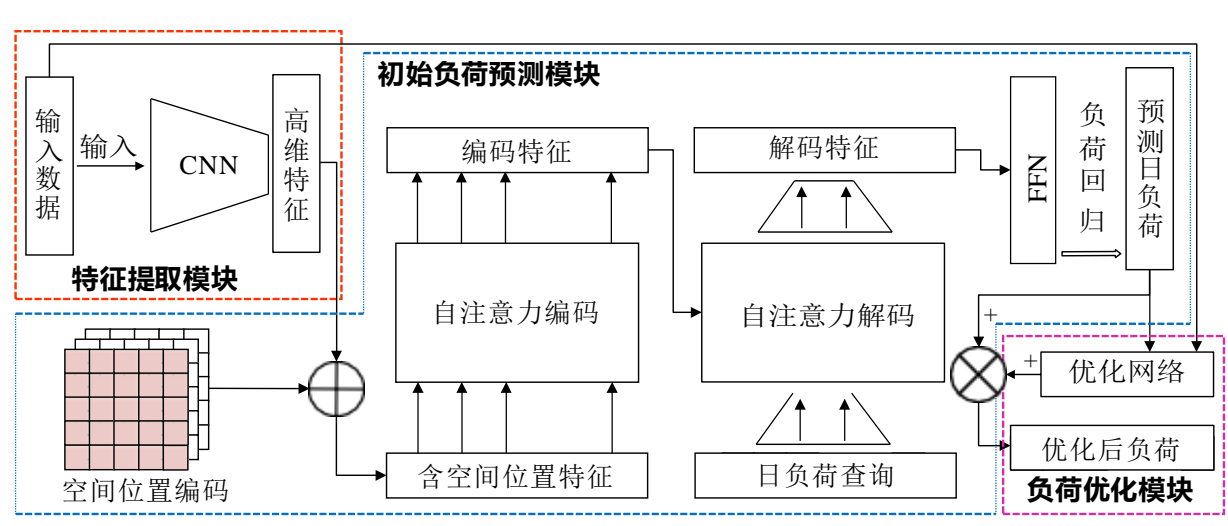

图 1 模型的整体结构

Fig.1 Overall architecture of model

## 1.2 特征提取模块

在负荷预测中，卷积神经网络常用于挖掘数据之间的局部相关性，获取高维特征向量。然而，之前的方法[10-13]所设计的卷积神经网络通常很浅，难以获取较为高级的语义信息。因此，本文设计了一个较深的卷积神经网络作为特征提取模块，用于挖掘语义信息，提高预测精度，具体结构如附录 A 图 A1 所示。

特征提取模块共包含 7 层卷积层，卷积核大小为 3×3，每一层卷积后都跟随一个批归一化层和一个激活层。除了第二个卷积层采用步长为 2 的卷积操作之外，其余步长均为 1。此外，模型还引入了残差连接，以防止梯度消失并加快模型收敛[15]。

## 1.3 初始负荷预测模块

初始负荷预测模块是 CNN-SAEDN-Res 中的核心部分，其性能直接影响模型的预测精度。在先前的预测方法中，不少研究人员使用序列模型处理数据和预测负荷。然而，在混杂了非序列因素的数据中，序列模型难以有效地建立前后信息之间的联系，导致预测精度下降。为了避免此问题，本文在初始负荷预测模块中舍弃了序列模型，而是采用自注意力编码解码网络对特征编码和解码，并通过前馈神经网络[16]实现负荷回归。

具体而言，初始负荷预测模块如附录 A 图 A2 所示，由空间位置编码、多个自注意力编码解码和前馈神经网络组成。

### 1.3.1 空间位置编码

序列模型的独特结构使其能够捕捉数据之间的序列关系，但自注意力编码解码网络无法自主获取数据的位置信息，使得特定时间与具有特定属性的负荷数据难以关联（负荷峰谷）。因此，需要进行位置编码（Position encoding, PE），以确保数据位置的严格对应。具体计算公式如下：

$$\begin{cases} V_{PE}\left(p,2r\right)=\sin\dfrac{p}{10\,000^{2r/d_{model}}} \\ V_{PE}\left(p,2r+1\right)=\cos\dfrac{p}{10\,000^{2r/d_{model}}} \end{cases} \tag{1}$$

式中：$V_{PE}\left(p,2r\right)$ 为 $\boldsymbol{V}_{PE}$ 在 $\left(p,2r\right)$ 处的元素值，$p$ 为数据在序列中的位置， $r$ 为向量维度的索引，$\boldsymbol{V}_{PE}$ 为位置编码矩阵；$d_{model}$ 为特征最大长度。

由于特征提取模块的输出为二维特征而非序列结构，因此需要对两个方向分别进行一次位置编码。编码完成后，将位置信息嵌入到输出特征中，以生成包含空间位置信息的高维特征，这些特征可作为自注意力编码模块的输入。

### 1.3.2 自注意力机制

在电力负荷预测任务中，注意力机制被用于关注重要信息和特征[17-18]。然而，注意力机制只关注因素信息与预测负荷之间的关系，起到的作用是局部的，忽视了各因素内部、各因素信息之间的关键信息，这对于包含时间信息的负荷预测任务来说并不友好。这是由于负荷变化通常具有一定的平滑属性，其趋势可以通过邻近负荷来反映。因此，为了建立数据之间的相关性，需要构造自注意力模型，并通过多次并行运行，构造多个自注意力，形成多头自注意力层。

自注意力机制具体流程如附录 A 图 A3 所示。首先，特征向量 $\boldsymbol{X}$ 与矩阵系数 $\boldsymbol{W}^{q}$ 、$\boldsymbol{W}^{k}$ 、$\boldsymbol{W}^{v}$ 计算分别得到矩阵 $\boldsymbol{Q}$ 、$\boldsymbol{K}$ 、$\boldsymbol{V}$ 。具体计算公式为：

$$\begin{cases} \boldsymbol{Q}=\boldsymbol{W}^{q}\boldsymbol{X} \\ \boldsymbol{K}=\boldsymbol{W}^{k}\boldsymbol{X} \\ \boldsymbol{V}=\boldsymbol{W}^{v}\boldsymbol{X} \end{cases} \tag{2}$$

然后将 $\boldsymbol{K}$ 转置后与 $\boldsymbol{Q}$ 相乘，计算得到特征相关性，再通过softmax函数对特征相关性归一化，得到注意力权重 $\alpha$ 。但为了防止 $\boldsymbol{K}^{T}\boldsymbol{Q}$ 的结果过大，

导致softmax后的偏导数过小，梯度趋近于 0，不利于模型收敛。因此，在进行归一化之前进行尺度变化。上述计算表达公式为：

$$\alpha = \mathrm{softmax}\left(\frac{\boldsymbol{K}^{\mathrm{T}}\boldsymbol{Q}}{\sqrt{d_{\mathrm{k}}}}\right) \tag{3}$$

式中：$d_{\mathrm{k}}$ 为矩阵 $\boldsymbol{K}$ 的列数。

最后，根据式(4)，将 $\boldsymbol{V}$ 与 $\alpha$ 计算得到自注意力 $\boldsymbol{A}$：

$$\boldsymbol{A} = \boldsymbol{V}\alpha \tag{4}$$

分析上述计算过程，自注意力机制通过输入向 量与自身转置相乘，获取输入向量在自身上的投影。投影的值越大，表明数据的关联程度越高。通过这样的方式，可以不受数据类型的限制，即使数据中混入了非序列因素，模型依然有能力通过全局信息 的作用关系选择重要的信息特征，遗忘不重要的部 分。

1.4 初始负荷优化模块

残差是指实际值与理论计算值之间的绝对误差，其产生原因多种多样，如星期类型的改变、气象因素的骤变等，都有可能导致模型预测不准，产生较大的残差值。但归结其本质原因，还是模型的学习能力不足，难以有效拟合日负荷曲线。

针对该问题，本文在初始负荷预测结束后添加负荷优化模块生成残差值，优化初始预测负荷的残差分布，从而改善第二次的负荷预测结果。该方法数学原理如下所示：

假设 $y_{\mathrm{Init}}$、$y_{\mathrm{Refine}}$、$e^*$ 分别为初始负荷预测结果、第二次负荷预测结果、负荷优化模块生成的残差值，那么三者之间应满足：

$$e^* = y_{\mathrm{Refine}} - y_{\mathrm{Init}} \tag{5}$$

又由于预测值 $y$、实际值 $\hat{y}$、残差 $e$ 之间满足：

$$y = \hat{y} - e \tag{6}$$

联立式(5)和(6)，可得：

$$e^* = e_{\mathrm{Init}} - e_{\mathrm{Refine}} \tag{7}$$

式(7)表明，负荷优化模块学习到的是一个残差值，旨在优化第二次负荷预测的残差结果。

本文所设计的负荷优化模块如附录A 图A4 所示，与特征提取模块所不同的是，在负荷优化模块中并不将输入特征当作二维结构利用卷积进行处理，而是利用了特征的序列结构，采用 GRU，紧跟一组前馈神经网络，生成负荷残差值，优化初始负 荷 。 这是 因 为 若 在 优 化 模 块 中 再采用 CNN-SAEDN 结构，会大大增加模型的时间、空间复杂度，降低优化模块性价比。

## 2 仿真设置

2.1 仿真流程

本文所提出的短期负荷预测方法工作流程如附录 A 图 A5 所示，具体步骤如下：首先，进行数据预处理，构建负荷预测输入特征集。然后将特征集划分为训练样本和测试样本，并将训练样本送入预测模型中，设置训练超参数开始训练，根据损失函数计算训练损失。每轮训练结束后，都要判断训练轮数是否达到预设值。若未达到预设值，则继续训练，并用优化器算法更新模型权重。若达到预设值，则停止训练，将测试样本送入到预测模型中，计算模型性能并预测负荷。

2.2 数据预处理

2.2.1 异常数据处理

电力负荷数据会受到人为失误、设备故障等多种因素的影响，导致数据出现异常值，这些异常值会影响模型的处理，进而影响预测效果。因此，本文对这些异常数据进行筛选、删除、替换，具体方法如下：

假设某日负荷数据为 $\{x_1, x_2, \cdots, x_h\}$，$x_h$ 为第 $h$ 次采集的负荷数据，$h = 24$。

由于异常数据属于离群点，会加大数据整体的离散程度，而标准差刚好可以反映这一特性。因此，本文通过计算标准差判断数据集合中是否存在异常数据。标准差 $\sigma$ 的计算公式为：

$$\sigma = \sqrt{\frac{1}{h}\sum_{a=1}^{n}\left(x_a - \mu\right)^2} \tag{8}$$

$$\mu = \frac{1}{h}\sum_{a=1}^{h} x_a \tag{9}$$

式中：$\mu$ 为样本均值，若 $\sigma > 140$（该数据通过 10 组异常数据和随机抽取的 100 组正常数据观察得出），则认为该日负荷数据集合离散程度过高，存在异常数据。但是，此时依然无法定位异常数据的具体位置。

由于日负荷数据的分布不服从正态分布，因此需要手工设定阈值，通过数据远离平均距离的标准差倍数来判断是否为异常数据。数据 $x_a$ 远离平均距离的标准差倍数 $m_a$ 的计算公式为：

$$m_a = \frac{\left|x_a - \mu\right|}{\sigma} \tag{10}$$

若 $m_a > 1$，则表明数据 $x_a$ 为异常数据，再求取异常数据 $x_a$ 左右正常数据的均值对 $x_a$ 进行填充替换。

2.2.2 数据归一化

数据归一化将量纲不同且数值差异较大的数据映射到[0,1]之间，能够加快模型收敛，减小预测误差。归一化计算公式为：

$$x^* = \frac{x - x_{\min}}{x_{\max} - x_{\min}} \tag{11}$$

式中：$x^*$为归一化后的数据；$x$ 为归一化前数据；$x_{\max}$、$x_{\min}$ 分别为一组数据中的最大值和最小值。

2.2.3 构建负荷预测输入特征集

构建合理的数据结构有助于模型建立数据之间的关系，由于电力负荷受到多重因素（如日期、时间、气象)的影响，因此需要综合考虑这些因素，构建多维特征向量作为模型输入特征，具体如附录 A 表 A1 所示。

在上述影响因素中，对于日期因素（除时间外），全部采用独热编码重新表示，便于模型进行数据处理。最终特征向量的空间维度为 9×24，其中，由于日期因素的范围为 1~31，用两行向量表示其独热编码。如附录 A 图 A6 所示，展示了 3.1 节算例分析所用数据集中某一日的输入特征。

2.3 仿真环境与超参数设置

本文所提出的负荷预测模型分为两部分训练，第一部分为基本预测网络 CNN-SAEDN， 使用 Adam 优化器，批样本数设置为 32，初始学习率设置为 0.001，在第 150 轮和第 300 轮时学习率较之前均减半，共训练 500 轮。第二部分为负荷优化网络，使用 Adam 优化器，批样本数设置为 16，初始学习率设置为 0.01，在第100 轮和第 200 轮时学习率较之前均减半，共训练 300 轮。为了证明本文所提负荷预测方法的有效性，所有模型基于网格搜索 法调整超参数，以便模型的性能均能达到近似最优。此外，所有仿真均在单张 GTX850M 上进行，编程语言为 python3.7，深度学习框架为 Pytorch，保证模型预测精度和速度对比的公平性。

2.4 损失函数

在本文中，采用具有易求导、存在解析解等优点的均方损失函数计算损失，函数表达式为：

$$F_{\text{loss}} = \frac{1}{b}\sum_{j=1}^{b} F_j \tag{12}$$

$$F_j = \frac{1}{n}\sum_{i=1}^{n}\left(\hat{y}_i - y_i\right)^2 \tag{13}$$

式中：$F_{\text{loss}}$ 为批样本数$b$ 下的总训练损失；$F_j$ 为第 $j$ 个样本下的日负荷均方损失；$n$ 为日负荷实际考核点数；$\hat{y}_i$ 为时间$i$ 的实际负荷；$y_i$ 为时间$i$ 的预测负荷。

之所以计算日负荷损失，是由于本文所提模型所采用的预测策略是多对多预测，即通过一次预测直接输出次日 24 小时的负荷预测结果。

2.5 性能指标

依据电力行业负荷预测技术规范[19]，选择日平均负荷预测准确率 $A_{\text{d}}$ 反映模型预测精度，日平均误差 $E_{\text{d}}$ 反映预测的整体性偏差，计算公式为：

$$A_{\text{d}} = \left[1 - \sqrt{\frac{1}{n}\sum_{i=1}^{n}\left(\frac{y_i - \hat{y}_i}{\hat{y}_i}\right)^2}\right] \times 100\% \tag{14}$$

$$E_{\text{d}} = \frac{1}{n}\sum_{i=1}^{n}\left|\hat{y}_i - y_i\right| \tag{15}$$

## 3 算例分析

3.1 GEFCom 数据集算例分析

为了验证模型的科学性与有效性，本文采用美国公共事业公司 2004 年 1 月 1 日至 2008 年 6 月 29 日的每小时负荷数据进行仿真[20]。在该数据集中，除负荷信息外，还包含日期、时间、节假日，气温信息。

在本文中，以 2004 年 1 月 1 日至 2008 年 6 月 22 日的数据作为训练样本集（39216 条样本)，对 2008 年 6 月 23 日至 2008 年 6 月 29 日的负荷进行日前预测（提前一天直接预测次日 24 小时的负荷变化)，并以 2.5 节中所提到的指标对模型进行量化分析。为了让对比更加全面，本文将预测结果分为四个部分进行比较，分别为一周中的第一天预测结果、五天工作日平均预测结果、两天休息日平均预测结果和一周平均预测结果，并辅以曲线图，将模型性能可视化后比较。

3.1.1 模型消融仿真结果

本文所提负荷预测模型主要由特征提取模块、初始负荷预测模块和负荷优化模块组成。因此，在该仿真中，通过消融模块分别展现各部分对预测精度的提升效果。分析表 1，添加初始负荷预测模块后预测准确率提高了 0.91%，平均误差降低了 2.92 kW，表明在该结构下的模型预测结果更加准确、

平滑。

表 1 本文方法在美国公开数据集上的消融仿真结果Table 1Ablation simulation results of proposed method onpublic datasets in the United States

| 模型设置 | | | 模型评估 | |
| --- | --- | --- | --- | --- |
| 特征提取模块 | 初始负荷预测模块 | 负荷优化模块 | $A_d$/% | $E_d$/kW |
| √ | | | 95.36 | 17.17 |
| √ | √ | | 96.27 | 14.25 |
| √ | √ | √ | **96.37** | **13.72** |

标注：√表示模型中含有该部分

在前文中提到，负荷优化模块的目的是为了学习残差，优化第二次负荷预测的残差结果，提高模型精度。为了验证该结论，本文对两次预测负荷的残差结果进行正态性检验[21]。

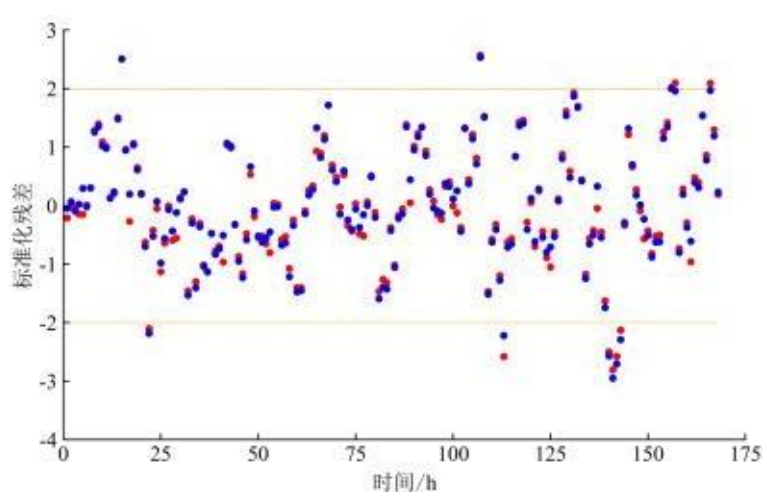

图 2 预测结果优化前后的残差标准化分布
Fig.2 Standardized distribution of residuals before and after optimization of forecasting results

如图 2 所示，以两种颜色的点分别展示两次预测负荷的残差标准化分布，其中红色为初始预测结果，蓝色为优化后的结果。统计发现，红色的点有 11 个落在两个标准差范围之外，而蓝色的点仅有 8 个。根据正态分布的理论，实验点的标准化残差落在范围之外的概率应小于 5%，即在本实验中异常点的个数应小于 8.4（168×5%）。初始负荷的预测结果未通过残差的正态性检验，模型对日负荷曲线的拟合能力不足，而优化后的结果则满足了这一检验要求。结合表 1，添加负荷优化模块后模型预测准确率提高了 0.1%，平均误差降低了 0.53kW，得到了小幅提升。上述结果表明，负荷优化模块优化了负荷的残差分布，改善了模型的拟合结果，提高了预测精度。

3.1.2 对比仿真

3.1.2.1 仿真说明

本节设置了两组仿真，用于验证模型性能。第一组仿真是本文方法与LSTM、GRU、CNN-LSTM、CNN-GRU 对比，主要表明本文方法相较于序列模型具有更好的预测效果。第二组仿真是本文方法与空间注意力机制[22]、通道注意力机制[23]的对比，主要验证自注意力机制相比空间、通道注意力能取得更优的预测结果。

3.1.2.2 与序列模型的对比仿真

该 仿真中 ， 所有 模型 使用的 CNN 与 CNN-SAEDN 中所使用的结构相同，LSTM 和 GRU 均设置为 3 层，通过相同结构的前馈神经网络得到日负荷预测结果。LSTM、CNN-LSTM、GRU、CNN-GRU 的批样本数均设置为 16，初始学习率为 0.01，训练轮数和递减效率与 CNN-SAEDN 一致。

仿真结果如附录 A 表 A2 所示，本文所提方法相较于 CNN、LSTM、CNN-LSTM、GRU、CNN-GRU 五种方法，在工作日时段，其准确率分别提高了 0.93%、1.23%、1.11%、1.00%、0.72%，平均误差分别降低了 3.28、5.43、4.11、3.12、3.20 kW。在休息日时段，所提方法准确率和平均误差依然取得最优的预测结果。综合一周的平均值进行分析，本文所提方法预测准确率和平均误差优势明显，表明模型的预测精度和稳定性优于其他方法。

除了进行数据对比外，本文还从日负荷曲线图 上对模型进行分析，本文提出的目的是为了在含有 非时序因素的数据中利用数据之间的内在联系，保 留重要信息，遗忘非重要信息，对预测结果起到积极的推动作用。观察图 3，发现除了文本所提方法之外，其他方法所预测的负荷高峰时间均向后偏移， 说明通过自注意力机制有效挖掘了负荷高峰与气 温、时间之间的联系，克服了序列模型无法有效处理非时序因素而导致的预测滞后问题。

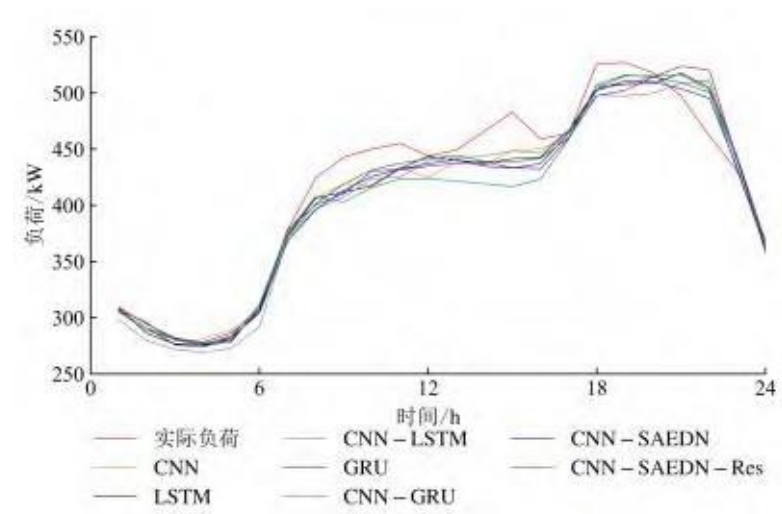

图 3 模型的周一负荷预测结果
Fig.3 Load forecasting results of model on Monday

为 了 让 结 论 更 具 说 服 力 ， 本 文 展 示 了 CNN-SAEDN 和序列模型的周负荷曲线图，如附录 A 图 A7-A11 所示。观察可知，序列模型易发生滞后现象，尤其是负荷高峰较为单一的工作日，该问题更加明显。

3.1.2.3 与注意力机制的对比仿真

该仿真中，所有模型均是在 3.1.2.2 节仿真中

CNN、CNN-LSTM、CNN-GRU 的基础上添加空间注意力（spitial attention, SA）和通道注意力（channel attention, CA），注意力模块被添加在卷积神经网络的残差模块中，除了初始学习率为 0.001 外[24]，模型训练的超参数与 3.1.2.2 节仿真相一致。

仿真结果如附录A 表 A3 所示，在结果对比中，CNN-SAEDN-Res 剔除了优化模块，仅用CNN-SAEDN 的预测结果与注意力机制下的各模型结果进行比较，突出自注意力机制相比较其他注意力机制的优势。

在含有空间注意力和通道注意力的模型中，根据一周的预测结果判断，表现最好的分别是基于 CNN-LSTM 的空间注意力 CNN-LSTM-SA 和基于 CNN-GRU 的通道注意力 CNN-GRU-CA。与本文所提方法进行综合性能比较，在工作日中，本文方法相较于 CNN-LSTM-SA、CNN-GRU-CA，其预测准确率分别提高了 0.55%、0.73%。综合分析一周的预测结果，本文方法的准确率分别提高了0.47%、0.63%，平均误差也有所降低，本文所提模型性能明显优于两者，表明在负荷预测任务中，数据之间的全局注意力所构造的全局相关性，相对于数据与预测结果之间的局部相关性，更能挖掘重要的数据特征，建立与预测结果之间的非线性关系[25]。

3.2 国内地区算例分析

本算例所采用的是国内南方某地区数据，该数据集提供了2016 年1月1日-2018年12月31日的电力负荷数据，一天采集24次，采样间隔为1h。此外，还提供了当日的最高气温、最低气温、平均气温、相对湿度和降雨量。

在本文中以 2016 年 1 月 1 日-2018 年 12 月 24 日的数据作为训练样本集（26112 条样本），对 2018 年 12 月 25 日-2018 年 12 月 31 日的负荷进行日前预测。为了展现模型的泛化性能，模型沿用第 3.1 节算例分析中的参数设置，未做出调整。

如附录 A 表 A4 所示，展示了各模型在该数据集上一周的平均预测准确率和平均误差，以及各模型进行单次日前预测所消耗的时间。对比可知，本文所提模型的预测准确率最高且平均误差最小。但所提模型的计算速度远远慢于其他模型，造成该结果的原因是在自注意力编码解码网络中进行了多次并行运算，导致模型的参数和计算量过多[26]。

虽然本文所提模型的计算速度慢于其他模型，但运算时长依然控制在毫秒级别，进行单次日前负荷预测仅需19.1ms，完全满足负荷预测任务的实时性需求。

## 4 结论

针对序列模型不能有效处理含有非时序因素的负荷数据导致预测精度不足的问题，本文提出了一种基于 CNN-SAEDN-Res 的短期日负荷预测模型，通过仿真验证，得出结论如下。

1）自注意力机制可以不受数据类型限制，即使是包含非时序因素的数据，也能通过全局注意力建立各数据之间的联系，通过该联系可以动态的为数据分配权重，自适应的保留和遗忘信息，在非时序数据中也能起到序列模型在时序数据中类似的效果。

2）通过对残差的分析和正态性检验，说明负荷预测结果可以在残差层面进行分析和优化，而不是仅限于通过加强模型的设计，提升模型的非线性学习能力来提高预测质量。

3）残差机制不仅仅体现在残差处理上，还包括负荷预测结果的多阶段优化。仿真表明，通过增加优化步骤可以提高模型的预测精度和稳定性。

但本文方法尚存不足之处：通过计算速度的对比，表明本文所提模型参数过多、冗余，未来将考虑通过剪枝技术进行结构优化。

参考文献：

作者简介：

崔杨(1980—)，男，教授，博士研究生导师，主要研究方向为电力系统运行分析、新能源联网发电关键技术等（**E-mail**：cuiyang0432@163.com）。

## Short-term power load forecasting method based on CNN-SAEDN-Res

CUI Yang[1]，ZHU Han[1]，WANG Yijian[1]，ZHANG Lu[2]，LI Yang[1]

(1. Key Laboratory of Modern Power System Simulation and Control & Renewable Energy Technology，Ministry of Education，Northeast Electric Power University，Jilin 132012，China；2. College of Information and Electrical Engineering，China Agricultural University，Beijing 100083，China)

**Abstract:**In deep learning, the load data with non-temporal factors are difficult to process by sequence models. This problem results in insufficient precision of the prediction. Therefore, a short-term load forecasting method based on convolutional neural network (CNN), self-attention encoder-decoder network (SAEDN) and resi- dual-refinement (Res) is proposed. In this method, feature extraction module is composed of a two-dimensional convolutional neural network, which is used to mine the local correlation between data and obtain high-dimensional data features. The initial load fore-casting module consists of a self-attention encoder-decoder network and a feedforward neural network (FFN). The module utilizes self-attention mechanisms to encode high-dimensional features. This operation can obtain the global correlation between data. Therefore, the model is able to retain important information based on the coupling relationship between the data in data mixed with non-time series factors. Then, self-attention decoding is per-formed and the feedforward neural network is used to regression initial load. This paper introduces the residual mechanism to build the load optimization module. The module generates residual load values to optimize the initial load. The simulation results show that the proposed load forecasting method has advantages in terms of prediction accuracy and prediction stability.

## 附录 A：

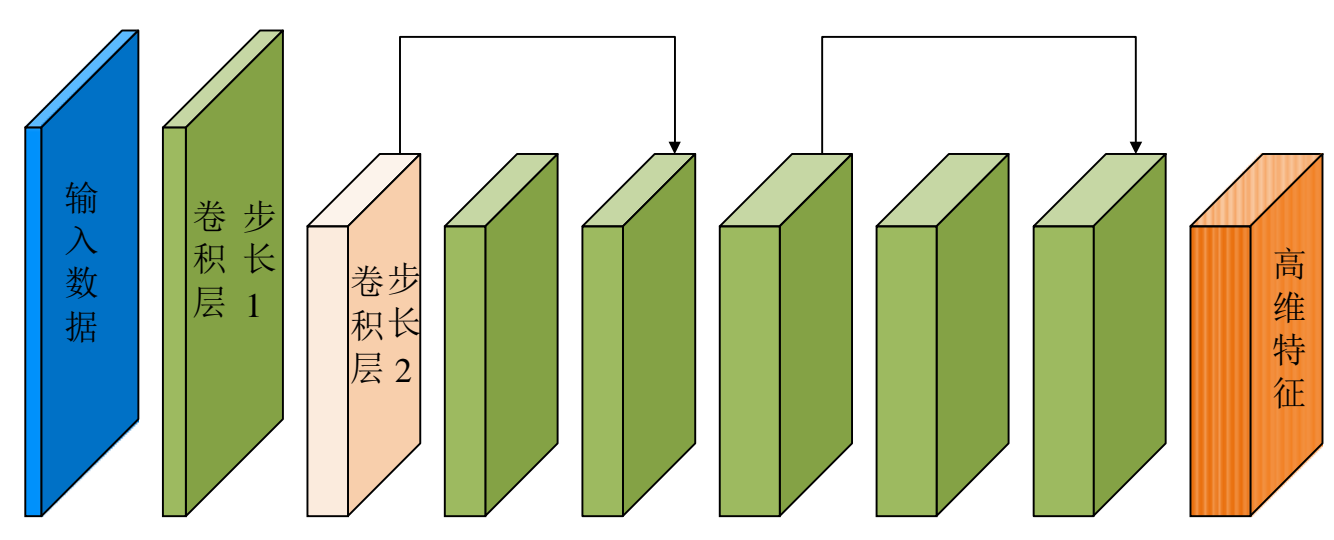

图 A1 特征提取模块结构

Fig.A1 Structure of feature extraction module

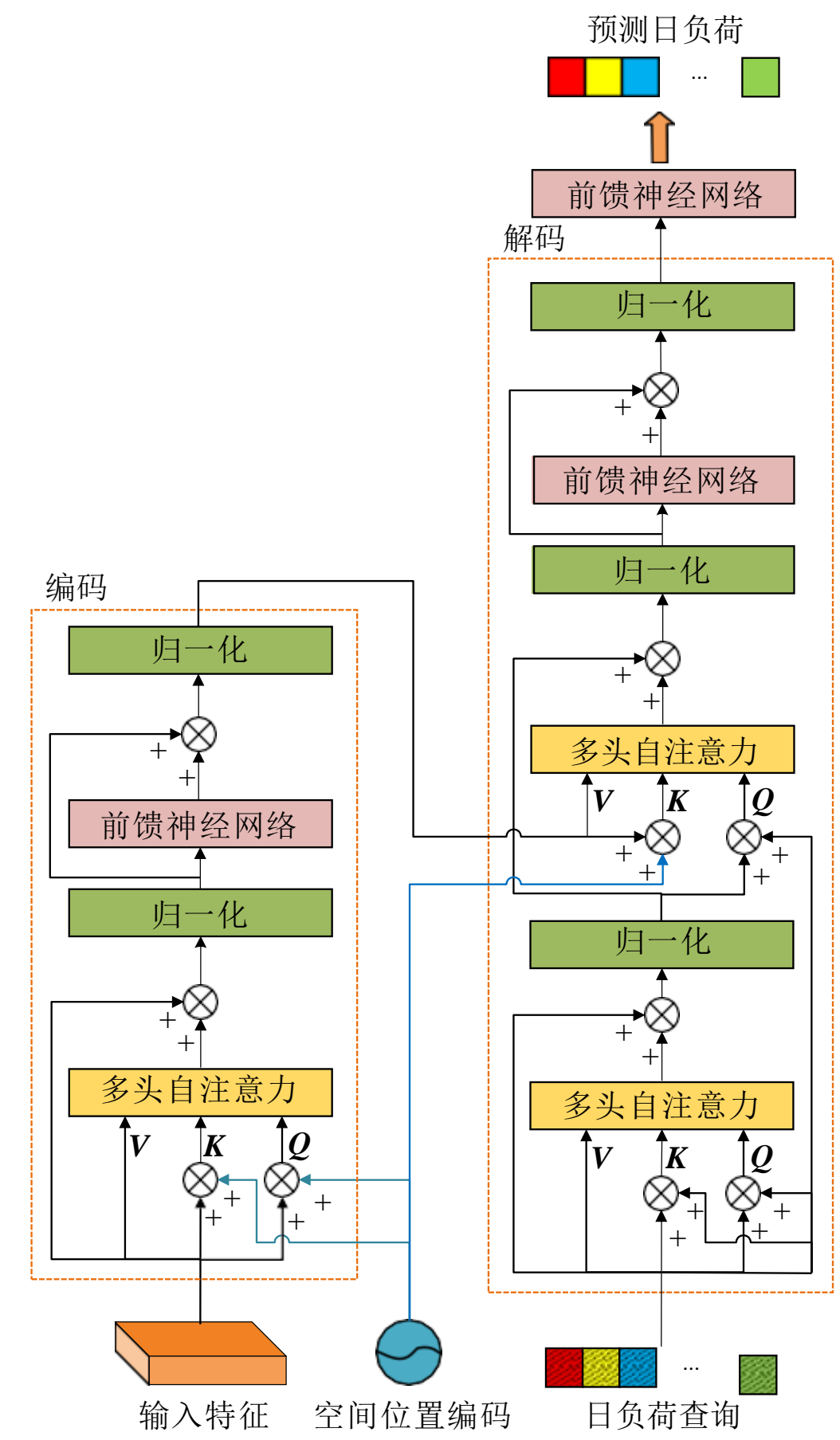

图 A2 初始负荷预测模块结构图

Fig.A2 Structure of initial load forecasting module

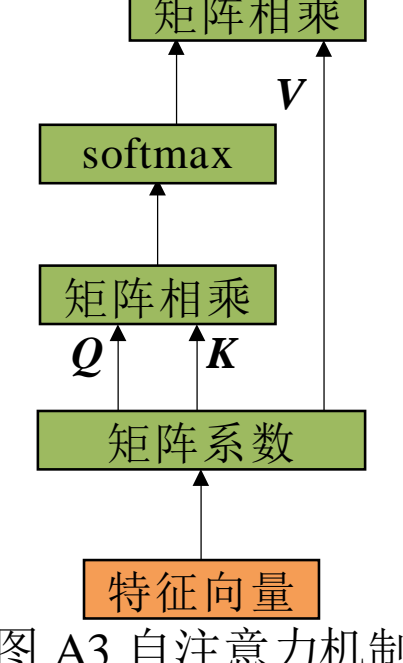

图 A3 自注意力机制

Fig.A3 Self-attention mechanism

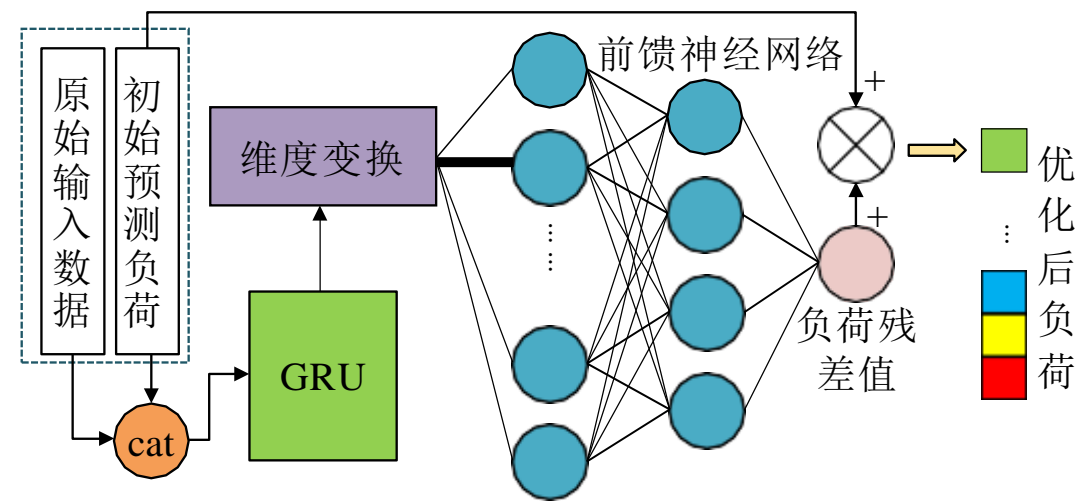

图 A4 负荷优化模块结构

Fig.A4 Structure of load refinement module

表 A1 输入特征集

Table A1 Input feature set

| 影响因素 | 因素类型 | 描述 |
| --- | --- | --- |
| 日期因素 | 月份 | 1—12 月 |
| | 日期 | 1—31 日 |
| | 周日期 | 一周 1—7 天 |
| | 时间 | 1—24 h |
| | 工作日/休息日 | 休息日包括周末和节假日 |
| 气象因素 | 气温/湿度/降雨量 | 24 h 气温/日最高、最低、平均气温/平均湿度、降雨量 |
| 相似负荷因素 | 前一个工作日/休息日负荷 | 根据数据集不同单位为 kW/MW |
| | 前二个工作日/休息日负荷 | |

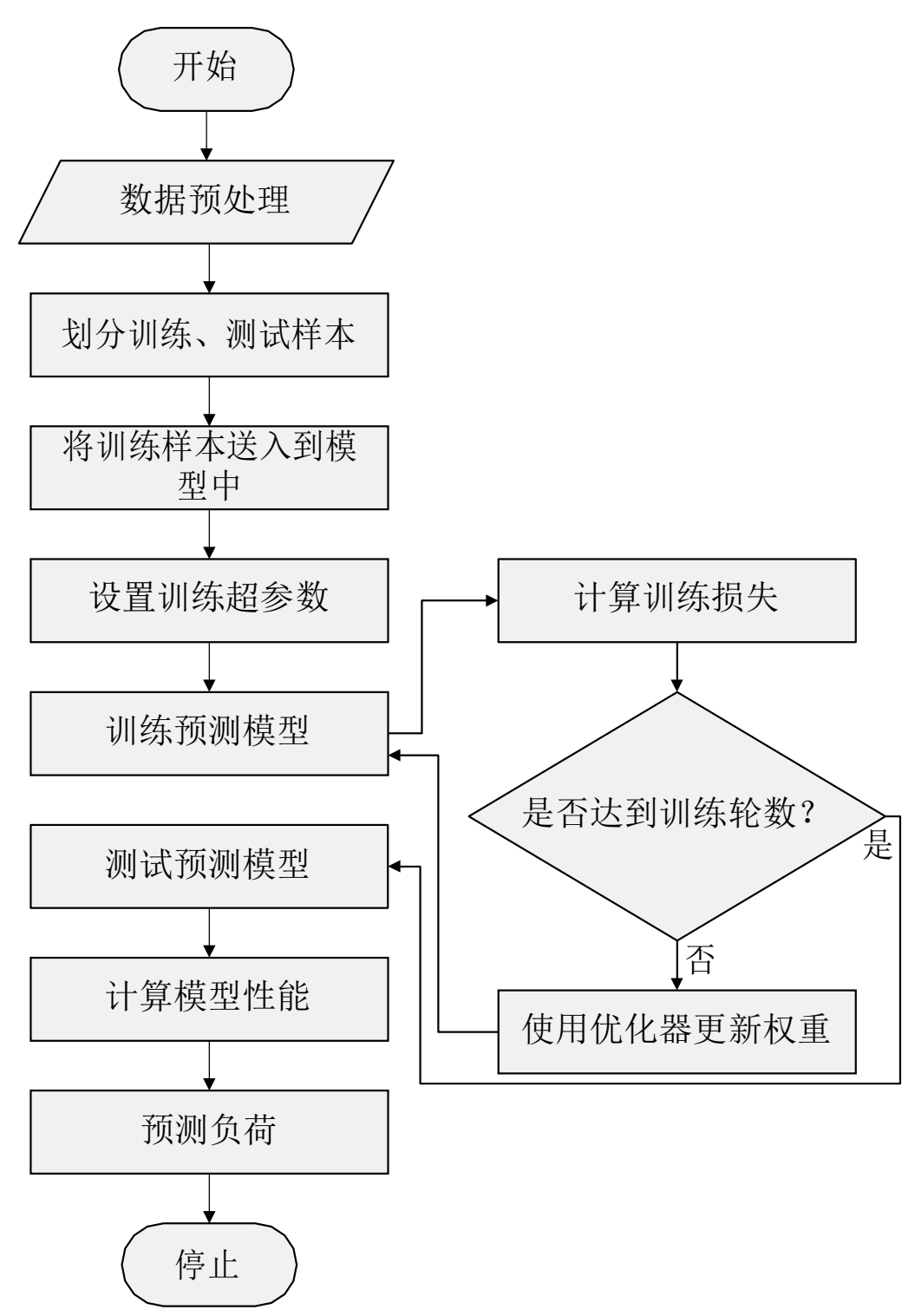

图 A5 短期负荷预测方法流程图

Fig.A5 Flowchart of short-term load forecasting method

| | | | | | | | | | | |
|---|---|---|---|---|---|---|---|---|---|---|
| 月份 | 1 | 0 | 0 | 0 | 0 | 0 | 0 | | 0 | 0 |
| 日期 | 0 | 0 | 1 | 0 | 0 | 0 | 0 | | 0 | 0 |
| | 0 | 0 | 0 | 0 | 0 | 0 | 0 | | 0 | 0 |
| 周日期 | 0 | 0 | 0 | 0 | 0 | 1 | 0 | | 0 | 0 |
| 休息日 | 0 | 1 | 0 | 0 | 0 | 0 | 0 | …… | 0 | 0 |
| 时间 | 1 | 2 | 3 | 4 | 5 | 6 | 7 | | 23 | 24 |
| 温度 | 48 | 48 | 47 | 49 | 49 | 49 | 49 | | 62 | 62 |
| 前一相似日负荷 | 461 | 438 | 447 | 439 | 448 | 485 | 534 | | 553 | 481 |
| 前二相似日负荷 | 484 | 457 | 450 | 448 | 444 | 490 | 523 | | 550 | 522 |

24

图 A6 模型输入特征

Fig.A6 Input feature of model

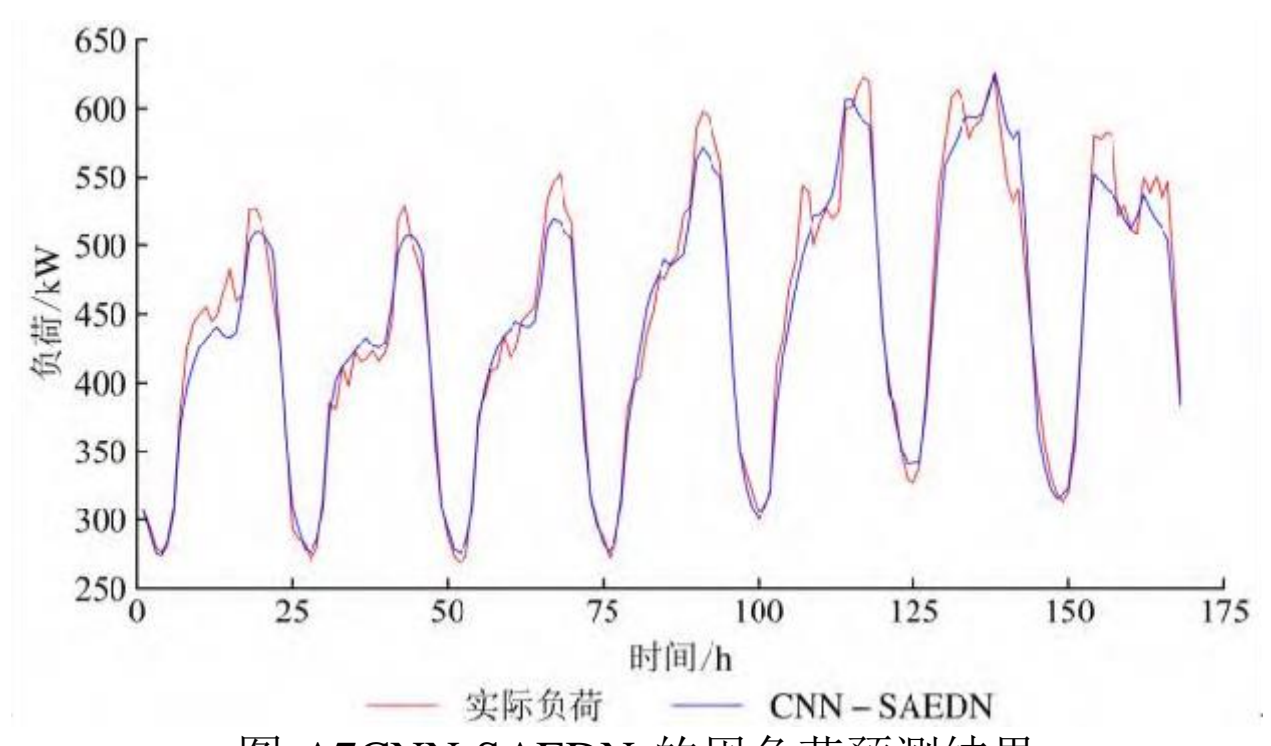

图 A7CNN-SAEDN 的周负荷预测结果

Fig.A7 Weekly load forecasting results of CNN-SAEDN

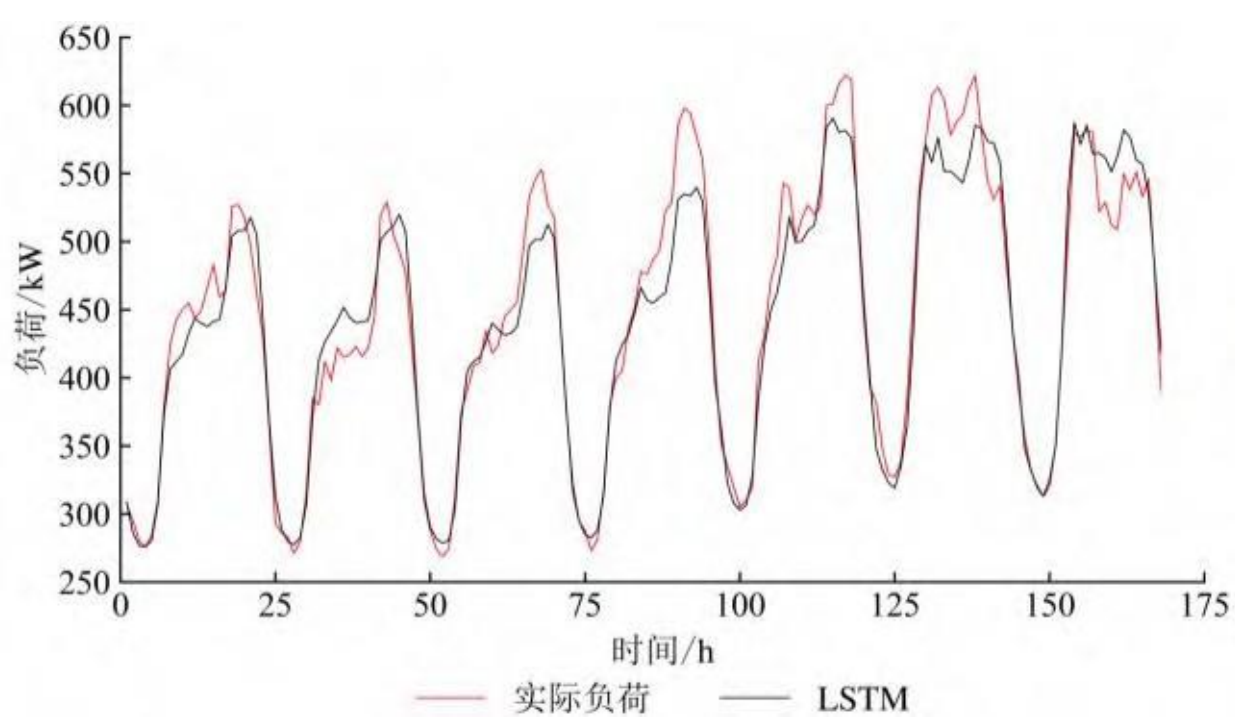

图 A8 LSTM 的周负荷预测结果

Fig.A8 Weekly load forecasting results of LSTM

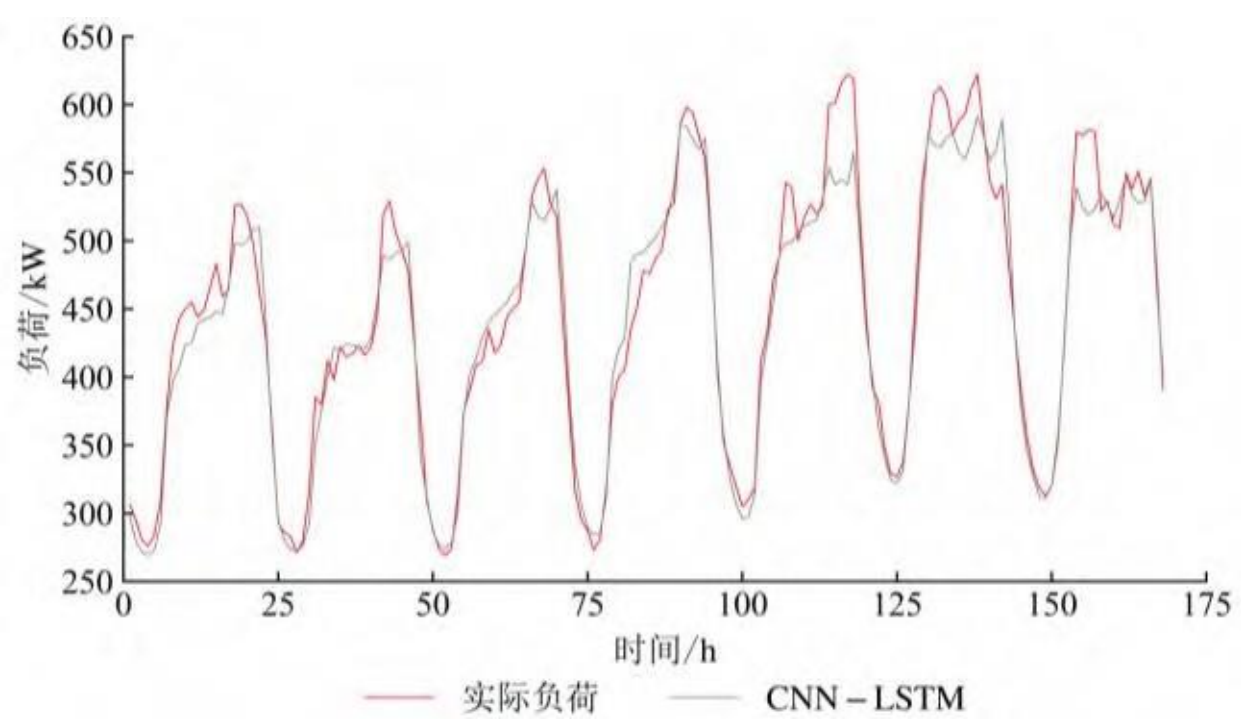

图 A9 CNN-LSTM 的周负荷预测结果

Fig.A9 Weekly load forecasting results of CNN-LSTM

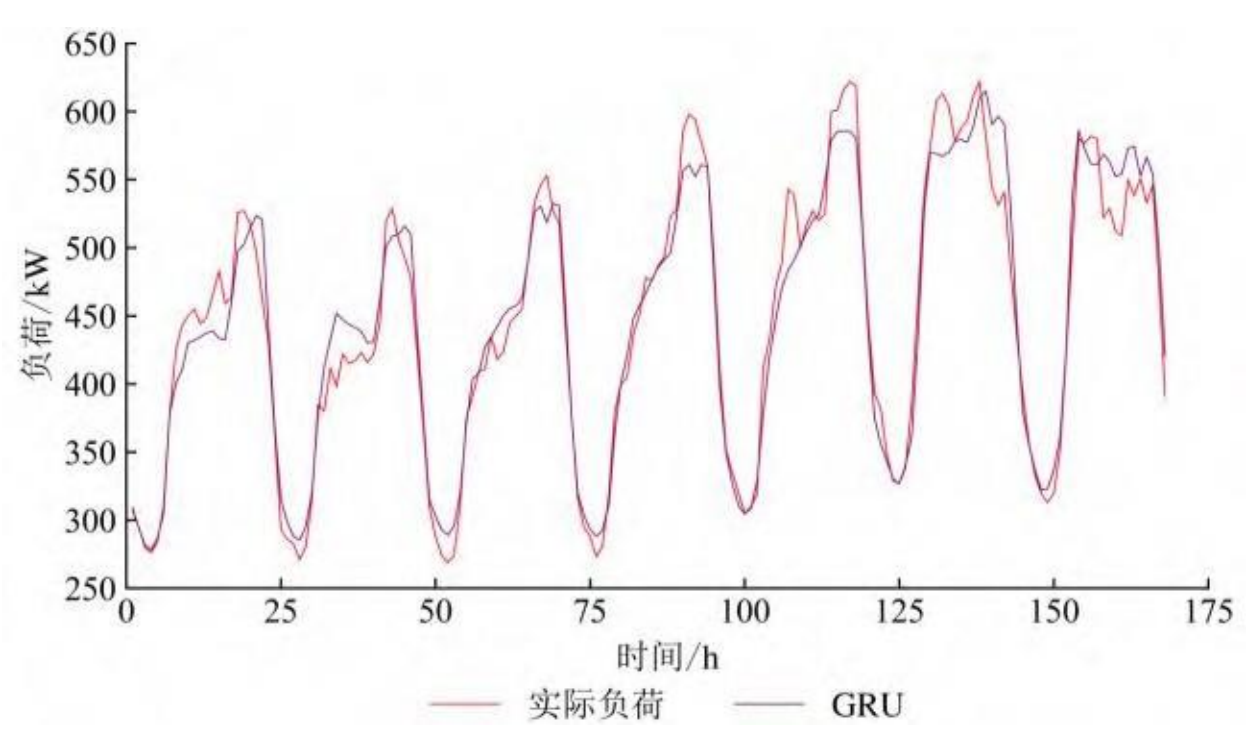

图 A10 GRU 的周负荷预测结果

Fig.A10 Weekly load forecasting results of GRU

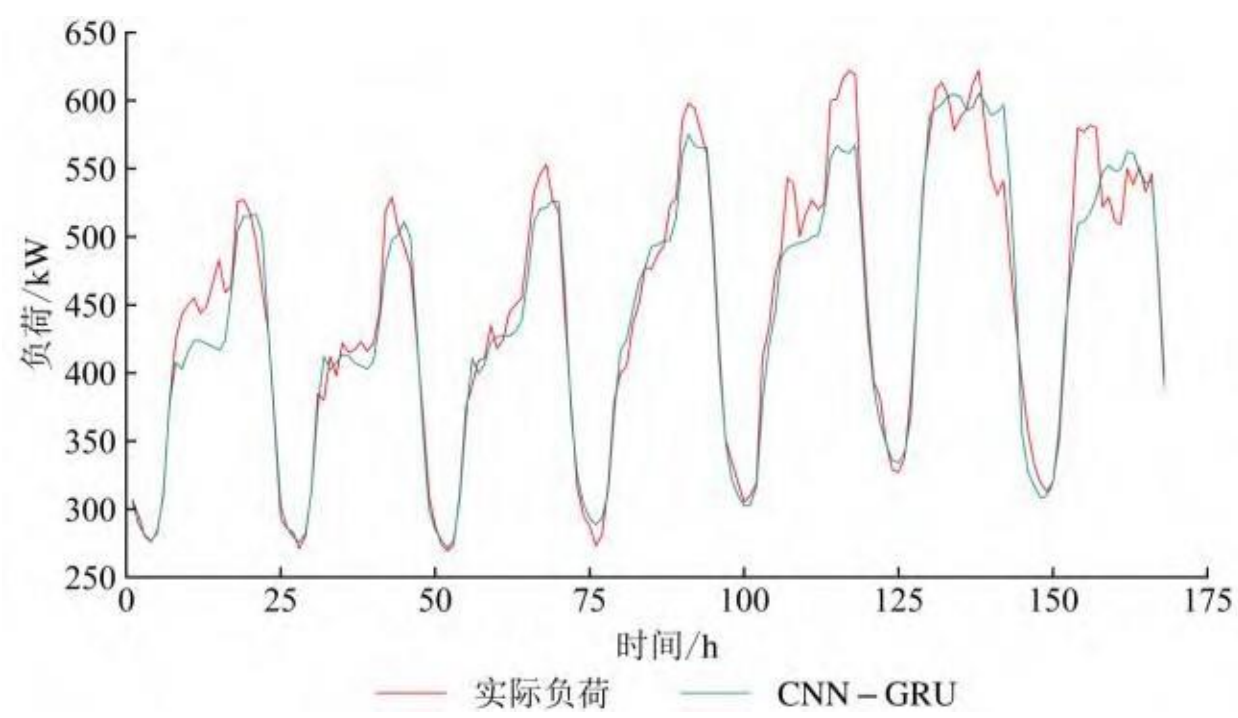

图 A11 CNN-GRU 的周负荷预测结果

Fig.A11 Weekly load forecasting results of CNN-GRU

表 A2 所提方法与序列模型的预测结果比较

Table A2 Comparison of forecasting results between proposed method and sequence model

| 方法 | 第一天 | | 工作日 | | 休息日 | | 一周 | |
|---|---|---|---|---|---|---|---|---|
| | $A_d$/% | $E_d$/kW | $A_d$/% | $E_d$/kW | $A_d$/% | $E_d$/kW | $A_d$/% | $E_d$/kW |
| CNN-SAEDN-Res | **96.28** | **12.32** | **96.58** | **12.30** | **95.82** | **17.30** | **96.37** | **13.72** |
| CNN-SAEDN | 95.79 | 14.42 | 96.49 | 12.89 | 95.73 | 17.65 | 96.27 | 14.25 |
| CNN | 95.89 | 14.27 | 95.65 | 15.58 | 94.64 | 21.16 | 95.36 | 17.17 |
| LSTM | 95.75 | 14.88 | 95.35 | 17.73 | 95.01 | 20.47 | 95.25 | 18.51 |
| CNN-LSTM | 95.14 | 17.98 | 95.47 | 16.41 | 95.57 | 17.55 | 95.50 | 16.74 |
| GRU | 95.14 | 16.54 | 95.58 | 15.42 | 94.79 | 21.05 | 95.35 | 17.03 |
| CNN-GRU | 94.48 | 18.28 | 95.86 | 15.50 | 94.46 | 21.41 | 95.46 | 17.19 |

表 A3 所提方法与注意力模型的预测结果比较

Table A3 Comparison of forecasting results between proposed method and attention model

| 方法 | 第一天 | | 工作日 | | 休息日 | | 一周 | |
|---|---|---|---|---|---|---|---|---|
| | $A_d$ /% | $E_d$ /kW | $A_d$ /% | $E_d$ /kW | $A_d$ /% | $E_d$ /kW | $A_d$ /% | $E_d$ /kW |
| CNN-SAEDN | 95.79 | 14.42 | **96.49** | **12.89** | **95.73** | 17.65 | **96.27** | **14.25** |
| CNN-SA | 95.84 | 13.15 | 95.65 | 14.88 | 94.56 | 20.86 | 95.34 | 16.59 |
| CNN-LSTM-SA | 95.92 | 14.63 | 95.94 | 14.05 | 95.45 | 19.44 | 95.80 | 15.59 |
| CNN-GRU-SA | 95.81 | 14.51 | 95.85 | 14.67 | 94.97 | 20.02 | 95.59 | 16.20 |
| CNN-CA | 95.79 | 12.75 | 95.41 | 15.84 | 95.00 | 18.40 | 95.30 | 16.57 |
| CNN-LSTM-CA | 95.17 | 15.97 | 95.76 | 15.21 | 95.34 | 17.64 | 95.64 | 15.90 |
| CNN-GRU-CA | 95.92 | 14.60 | 95.76 | 15.08 | 95.57 | 17.37 | 95.70 | 15.73 |

表 A4 我国南方某地区电力负荷数据集上的模型预测结果比较

Table A4 Comparison of model forecasting results on a power load dataset of a region in southern China

| 方法 | 一周 | | 计算速度/ms |
|---|---|---|---|
| | $A_d$ /% | $E_d$ /MW | |
| CNN-SAEDN-Res | 97.56 | 117.69 | 19.1 |
| CNN | 95.85 | 208.13 | 3.7 |
| LSTM | 95.76 | 224.37 | 2.9 |
| CNN-LSTM | 96.78 | 163.29 | 4.6 |
| GRU | 94.49 | 265.24 | 2.3 |
| CNN-GRU | 96.46 | 180.14 | 4.1 |
| CNN-SA | 96.37 | 190.94 | 5.9 |
| CNN-LSTM-SA | 97.15 | 143.46 | 7.1 |
| CNN-GRU-SA | 97.04 | 144.68 | 6.7 |
| CNN-CA | 96.30 | 181.61 | 5.3 |
| CNN-LSTM-CA | 97.19 | 141.98 | 6.3 |
| CNN-GRU-CA | 96.97 | 151.10 | 6.1 |